A Mathematical Model for Estimating Biological Damage Caused by Radiation

Yuichiro Manabe,

Kento Ichikawa[1],

Masako Bando[1,2]

Division of Sustainable Energy and Environmental Engineering, Graduate School of Engineering, Osaka University, 2-1 Yamada-oka, Suita, Osaka 565-0871, Japan

[1]Jein Institute of Fundamental Science, Venture Business Laboratory, Kyoto University, Yoshida, Honmachi, Sakyo-ku, Kyoto-shi, 606-8501, Japan

[2]Institute of Fundamental Physics, Kyoto University, Kitashirakawa oiwake-cho, Sakyo-ku, Kyoto-shi, 606-8502, Japan

Abstract

We propose a mathematical model for estimating biological damage caused by low-dose irradiation. We understand that the Linear Non Threshold (LNT) hypothesis is realized only in the case of no recovery effects. In order to treat the realistic living objects, our model takes into account various types of recovery as well as proliferation mechanism, which may change the resultant damage, especially for the case of lower dose rate irradiation. It turns out that the lower the radiation dose rate, the safer the irradiated system of living object (which is called symbolically "tissue" hereafter) can have chances to survive, which can reproduce the so-called dose and dose-rate effectiveness factor (DDREF).

# 1. Introduction

It has now become one of the serious problems how the low-dose radiation hurts biological objects. If it is merely a physical process, so a most reasonable hypothesis of the frequency of radiation-induced mutations is that it is proportional to total dose irradiation, which is usually called LNT hypothesis. However, this is nowadays adopted as an important basic assumption to estimate the low-dose radiation risk. Indeed, the first experiment of Drosophila spermatozoa by Hermann Joseph Muller[1] provided us with clear evidence which proves the LNT hypothesis.

However the above results were obtained only under a certain cell conditions. It is well known that spermatozoa of any kinds of living object are known to have no recovery system. Thus the question arises how the data of the effects is changed under the other situations than spermatozoa. Moreover it has now become more important question how the bulk of the radiation dose causes biological damage in living mankind. Also it is of both practical and fundamental importance to question whether the mutation rates or amount of radiation damage of human body may depend only on the total amount of radiation intensity itself or on the variation of the radiation dose exposure processes. So far as we adopt the LNT hypothesis the answer would be the former, namely the total amount of radiation dose does determine the total damage.

We have now accumulated data to answer such questions; First, there has been increasing evidences that induction of mutation may not be as direct an action as had often been supposed, and that the mutation process in the gene depends on the variation of its cellular environment[2].

The mutation rates induced by irradiation change if we take account of most of the mechanism working in living bodies. Since there are known many effects, recovery of DNA, apoptosis, bystander effect, radiation homeostasis. Those effects may be classified into two characteristics, enhancement on the one side and depression on the other side. The first one actually induces a kind of death of broken cells, caused by some apoptosis effects. The second one is somewhat related to so called recovery effect. In any case, those effects appear as a certain deviation from LNT to opposite direction, enhanced effect for the first case or depressive effects for the latter case.

In this paper, we propose a mathematical model to estimate the risk induced by irradiation in living bodies, which takes account of the above effects working in living bodies. Here we consider a system of cells which has certain function such as tissue or organ, with its capacity $K$, the maximum number of cells inside this system. Hereafter we call it symbolically "tissue". Suppose that at $t=0$ a tissue contains only normal cells with its number $N_0$ and is exposed by radiation with the rate $r(t)$. It is interesting to see the time dependence of the numbers of normal and broken cells therein. The risk estimate may be related to the number of broken cells which may turn to a cancer tumor. In our mathematical model, we define the asymmetry to give us a reference index of the risk.

In section 2 we introduce a simplest mathematical model and propose a derivative equation of motion and show the results of numerical calculation. Section 3 we include proliferation of cells to the simplest model. Then we introduce more realistic model for living objects by taking account of effects, repair and apoptosis in section 4, followed by numerical calculations for typical cases in section 5. Concluding remarks and future problems will be presented in section 6.

**2. Simple Model without Proliferation Function**

Here we consider a system of cells which has a certain function such as tissue or organ. Hereafter we call it symbolically "tissue". Suppose that at $t=0$ a tissue contains only normal cells with its number $N_0$ and is exposed by radiation with the rate $r(t)$. The risk estimate may be related to the number of broken cells which may turn to a cancer tumor.

Before going into biological objects let us see simple process where cells have neither recovery effects nor proliferation function. Let us denote the numbers of normal and broken cells, $N_n$ and $N_b$, respectively. The normal cells of a system are broken due to the irradiation strength. Namely the total number of normal cells of a tissue decreases according to the strength of irradiation, for which we here use the unit Gy. This represents the unit of absorbed dose, the absorption of one joule of energy, inducing ionizing radiation, per kilogram of matter. Thus in terms of Gy, the irradiation strength rate $r(t)$ deposits a corresponding increment of energy per time in unit volume of tissue, and thus the derivative of total number of broken cells, is proportional to the amount of irradiation strength rate $r(t)$. Therefore the numbers of normal and broken cells, obey the following differential equations;

$$\frac{d}{dt}N_n = -c \cdot r(t),$$
$$\frac{d}{dt}N_b = c \cdot r(t), \quad ^a$$
$$N_n \geq 0, \quad (2.1)$$

---

[a] In the low dose region, the effect of radiation yields mutation dominantly, and we here neglect the contribution of cell death. Thus we assume, for the moment, that the influence of radiation causes mutation only, not cell death. In this case, the total number of normal and broken cells, $N_n + N_b$, is constant $N_0$, as is seen in Eq.(2.1).

where $c$ is a breaking coefficient to the irradiation strength rate $r(t)$. This is what we call radiosensitivity. In general, the coefficient $c$ might be determined by radiation cross section of cells, cell density, and the related surrounding conditions. However, the unit Gy is defined as the absorption of one joule of energy, in the form of ionizing radiation, per kilogram of matter. Note that in this paper we are focusing on the case of low dose region and assume that $c$ is independent of $N_n$. Such kind of treatment may be similar to the situation in which nuclear physicists often employs the concept of nuclear matter which is defined as an idealized system consisting of a huge number of protons and neutrons with finite density. Thus $c$ is independent of cell density unless normal cell number is extremely small.

On the contrary if the cell density becomes very small and the number of normal cells is very small and the radioactive energy deposit is not fully poured into the breakdown of normal cells the ratio of which is in proportion to the number of normal cells. So more general form of the coefficient $c$ can be written as

$$c = \bar{c} g(N_n), \quad (2.2)$$

with

$$g(N_n) \xrightarrow[\beta N_n \ll 1]{} N_n, \quad (2.3)$$
$$\xrightarrow[\beta N_n \gg 1]{} 1. \quad (2.4)$$

As such examples we can take

$$g(N_n) = 1 - \exp(-\beta N_n), \quad (2.5)$$
$$g(N_n) = 1 - \frac{1}{1 + \beta N_n}. \quad (2.6)$$

We show the time dependence of, $N_n$ and $N_b$ for several cases of $\beta$ in Fig.1 using Eq. (2.5) as an example.

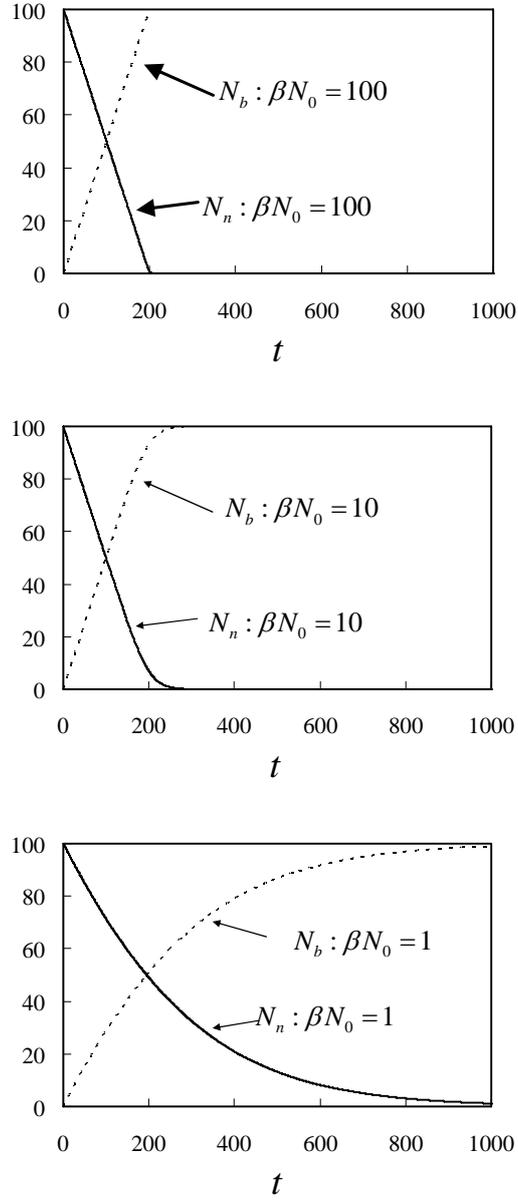

Fig. 1. Total dose dependence of the numbers of normal and broken cells in non-living objects, for three cases with different parameters $\beta = 1, 0.1, 0.01$ (see Eq.(2.5).) of the with initial condition $N_n(t=0) = N_0 = 100, N_b(t=0) = 0$ for irradiation strength $cr = 0.5$. The total number of normal and broken cells, $N_n(t) + N_b(t)$, is constant $N_0$.

Here we take the units of all the variables appropriately so that all the variables are made dimensionless except radiation dose. The solutions are straight forwardly obtained;

$$N_n(t) = N_0 - N_b(t),$$
$$N_b = c\int_0^t r(t')dt' \equiv c \cdot R(t), \qquad (2.7)$$

where $R(t)$ is the total amount of irradiation dose which the object is exposed during the time interval between $t=0$ and $t=t$. This shows what we call LNT as shown in Fig. 1. Note that the number of broken cells depends only on the total amount of exposure of radiation dose and is independent of the radiation strength. This dependence is actually observed in the Mueller's experiments on the physiological and genetic effects of radiation (X-ray mutagenesis). Note that spermatozoa which are used by Mueller's experiments are known to have no recovery function.

Let us consider the case where $c$ is time independent. Suppose that $r(t)$ is also time independent, namely $r(t) = r$, we have the following simple solutions,

$$N_n(t) = N_0 - c \cdot rt,$$
$$N_b = c \cdot rt. \qquad (2.8)$$

with initial conditions, $N_n(t=0) = N_0$, $N_b(t=0) = 0$. If we define the exposure dose time interval by $T$, the number of broken cells is proportional to $T$, accounting the total dose, $R = rT$.

## 3. Simple Model with Proliferation Function

Next let us introduce a simplest model with proliferation function.

Consider the case of the cells in a tissue, which are exposed by irradiation with intensity $r(t)$. Let $N_n$ and $N_b$ be the normal and broken cell numbers again, which continue to proliferate themselves with the proliferate rates, $\alpha_n$, $\alpha_b$ respectively,

$$\frac{d}{dt}N_n(t) = \left(1 - \frac{N_n(t)}{K}\right)\alpha_n N_n(t) - cr(t), \quad (3.1a)$$

$$\frac{d}{dt}N_b(t) = cr(t) + \alpha_b N_b(t), \quad (3.1b)$$

with irradiation rate $r(t)$ and $c$ is its breaking coefficient with the initial conditions at $t = 0$, $N_n(t=0) = N_0, N_b(t=0) = 0.$

Before discussing the general case, let us consider the case, $r = 0$. The solution of Eq. (3.1a) is

$$N_n(t) = \frac{K}{1 + \left(\frac{K}{N_0} - 1\right)\exp(-\alpha_n t)}. \quad (3.2)$$

This is so called logistic function which is commonly used for population growth. It shows the "S-shaped" curve with its slope tending to zero when it reaches its maximum $K$ which its capacity allows. Here we assume that the number of normal cells is controlled so as not to exceed the its carrying capacity $K$ and the proliferation rate tends to zero when $N_n$ approaches $K$. This is because the proliferation rate of normal cells is controlled by the suppression factor $\left(1 - \frac{N_n(t)}{K}\right)$.

In the case $r \neq 0$, the number of normal cells decreases due to the mutation caused by irradiation. If $r(t) = r$, which is independent of time $t$, Eq.(3.1a) can be rewritten as

$$\frac{d}{dt}N(t) = \alpha N(t)\left(1 - \frac{N(t)}{\mathrm{K}}\right),$$

$$N_n(t) \equiv N(t) + \beta,$$

$$\beta \equiv \frac{K}{2} - \frac{K}{2}\sqrt{1 - \frac{4cr}{\alpha_n K}},$$

$$\alpha \equiv \frac{\alpha_n}{K}(K - 2\beta) = \alpha_n\sqrt{1 - \frac{4cr}{\alpha_n K}},$$

$$\mathrm{K} \equiv K - 2\beta = K\sqrt{1 - \frac{4cr}{\alpha_n K}}. \quad (3.3)$$

This time, the solution of Eq. (3.3) is

$$N(t) = \frac{\mathrm{K}}{1 + \left(\dfrac{\mathrm{K}}{N(0)} - 1\right)\exp(-\alpha t)}. \quad (3.4)$$

More explicitly,

$$N_n(t) = \frac{K\sqrt{1 - \dfrac{4cr}{\alpha_n K}}}{1 + \left(\dfrac{K\sqrt{1 - \dfrac{4cr}{\alpha_n K}}}{N_0 - \beta} - 1\right)\exp\left(-\alpha_n\sqrt{1 - \dfrac{4cr}{\alpha_n K}}\,t\right)} + \beta,$$

$$\beta \equiv \frac{K}{2} - \frac{K}{2}\sqrt{1 - \frac{4cr}{\alpha_n K}},\ cr \leq \frac{\alpha_n K}{4} \quad , (3.5)$$

where $\beta$ must satisfy

$$\frac{\alpha}{K}\beta^2 - \alpha\beta + c = 0, \quad (3.6)$$

in order to reproduce the similar differential form as Eq.(3.1a). We have two solutions, one of which yields the maximum value of $N(t)$ larger than $K$.

which has two solutions, one of which yields the maximum value of $N(t)$ larger than

K.

Comparing Eq.(3.2) with Eq.(3.5), we see that $K$ and $\alpha$ are replaced by the $K$ and $\alpha_n$ with the factor $\sqrt{1-\dfrac{4cr}{\alpha_n K}}$.

For the broken cells, the solution of Eq.(3.1b) becomes,

$$N_b(t) = \frac{cr}{\alpha_b}\left(-1+\exp(\alpha_b t)\right), \quad (3.7)$$

for the case $r(t)=r$. Here we have seen the essential difference between normal and broken cells. While broken cells proliferate themselves without no control, the proliferation of normal cells tend to approach its maximum number.

From Eq.(3.2), we see in the number of normal cells, $N_n$ the so called "logistic curve" which is commonly observed in the growth of population, where in the initial stage of growth is approximately exponential approaching to some saturation region where the growth slows, and tends to its maximum value $K$. As for the case $cr \neq 0$, as seen Eq.(3.5), it approaches to the maximum value $\dfrac{K}{2}\left(1+\sqrt{1-\dfrac{4cr}{\alpha_n K}}\right)$ when $t$ becomes large, while $t \ll 1$,

$$N_n(t) \to \left[\left(1-\frac{N_0}{K}\right)\alpha_n N_0 - crt\right] + N_0. \quad (3.8)$$

On the other hand, $N_b$ behaves at the first stage, namely $t \ll 1$,

$$N_b = \frac{cr}{\alpha_b}\left(-1+\exp(\alpha_b t)\right) \to crt. \quad (3.9)$$

This shows linear dependence of the number of broken cells on its irradiation dose rate. After time interval T it becomes proportional to total dose rate $rT = R$. Thus, so far as the irradiation time is small enough, i.e., $\alpha_b T \ll 1$, the number of broken cells increases almost linearly on the total radiation dose $R$.

$$N_b = cR, \quad R = rT. \qquad (3.10)$$

Note that it depends only on the total dose $R$ independently of the proliferation rate $\alpha_b$. However, as seen from Eq.(3.2) the second term of $N_b(t)$ gradually dominates and then the number of broken cells gradually brows up exponentially. Fig.2 shows typical examples of time dependence of the numbers of normal and broken cells, $N_n$, $N_b$. Their behavior is consistent with our argument. Note that $cr$ becomes large, the number of normal cells becomes 0 at critical time, $t_c$, yielding the death of tissue. Even if $cr$ is less than the critical value of irradiation rate, $N_n$ seems to approach to $K$ but never becomes maximum number $K$ under the constant exposure of irradiation even if it is as low as natural radiation. This is inevitable so far as recovery effect is missing. We shall discuss such effect in the next section.

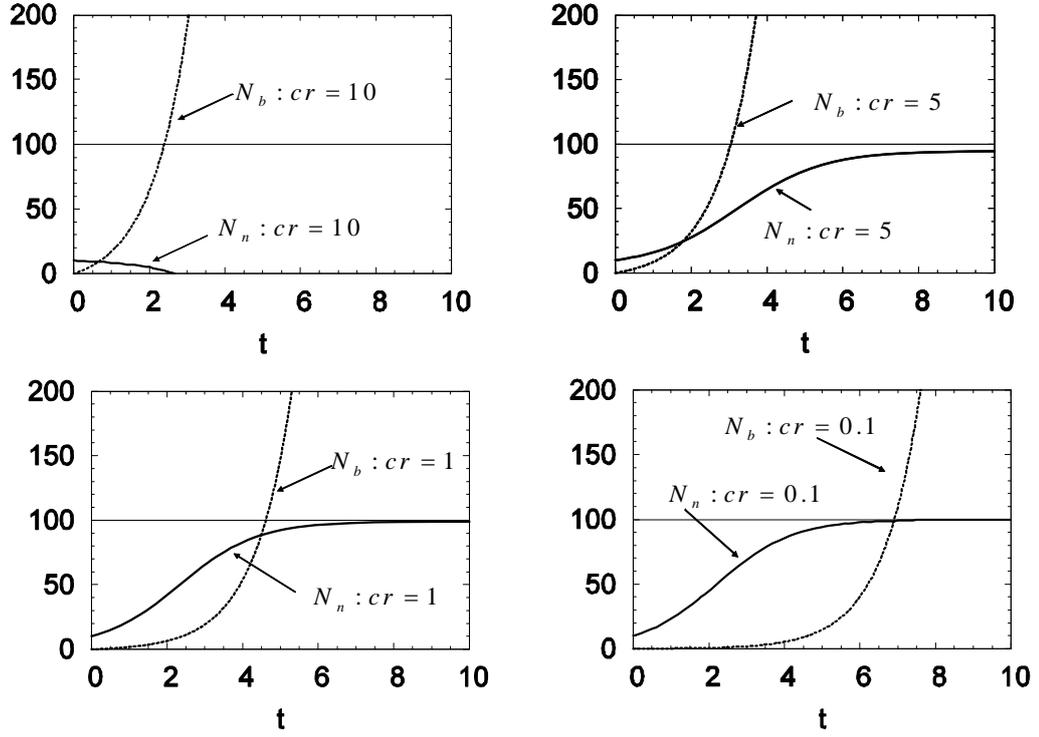

Fig. 2. Time dependence of $N_n$ and $N_b$ for various irradiation rate $r$. The vertical axis expresses number of cells and horizontal axis, time. Time dependence of $N_n$ corresponds to solid line, and $N_b$ dotted line, for different irradiation strengths $cr = 0.1, 1, 5, 10$ under the initial condition, $N_n(t=0)=N_0$, $N_b(t=0)=0$.

Next we define that the actual damage of a tissue occurs when the number of broken cells exceeds the one of normal cells. Here we introduce the asymmetry parameter $A$ as

$$A(t) = \frac{N_b(t) - N_n(t)}{N_b(t) + N_n(t)}, \quad (3.11)$$

which is in general also time dependent variable. If $A(t)$ becomes $0$, the cancer risk becomes appreciable and we define the tissue turns cancerous when $A$ becomes positive. In Fig.3 we demonstrate the time dependence of $A(t)$. $A(t)$ corresponds to the case of no recovery effect. So the number of broken cells increases as time development from the initial condition $A(0) = -1$ (only normal cells exists) and inevitably becomes positive at certain critical time interval. This tendency becomes more remarkable for larger irradiation dose $r$.

In this way it becomes more visually evident that in this simplest case the number of broken cells inevitably exceed $0$ and broken cells dominates the issue soon or later. The stronger the radiation strength, the critical time becomes shorter. Thus we cannot stop outbreak of proliferation of broken cells, and living body without any recovery effects is inevitably lead to cancer development.

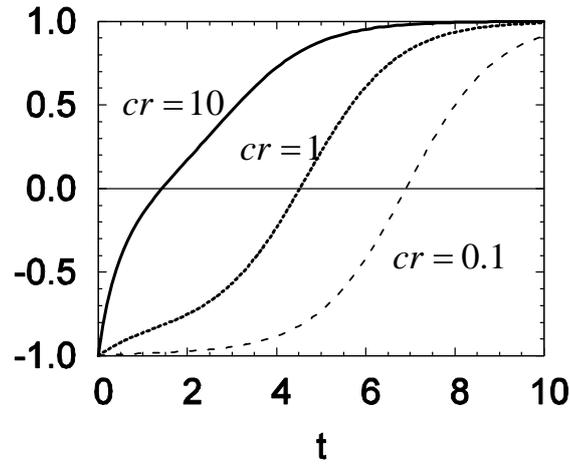

Fig. 3. Time dependence of $A(t) = \dfrac{N_b(t) - N_n(t)}{N_b(t) + N_n(t)}$, which may represent a sort of degree of canceration, for various irradiation strength $r$. Time dependence of $A(t)$ corresponds to solid line, dotted line and dashed line for different irradiation strength $cr = 10, 1, 0.1$, with the parameters $\alpha_b = \alpha_n = 1$. The initial condition is $N_n(t=0) = N_0$, $N_b(t=0) = 0$.

## 4. Model with Recovery Effects

Next we take account of restore and apoptosis effects against broken cells. Then the derivative equations for the numbers of normal and broken cells are expressed as,

$$\frac{d}{dt}N_n(t) = \left(1 - \frac{N_n(t)}{K}\right)\left(\alpha_n N_n(t) + \mu_r N_b(t)\right) - cr,$$

$$\frac{d}{dt}N_b(t) = cr(t) + \left(\alpha_b - \mu_r - \mu_a\right)N_b(t), \quad (4.1)$$

with $\mu_r$, $\mu_a$, the rates of inducing restore and apoptosis of broken cells, respectively. Again we assume that the suppression factor controls the number of normal cells so as not to exceed its maximal.

The solution of Eq. (4.1) is easily obtained if $r(t) = r$, time independent;

$$N_b(t) = \frac{cr}{\mu}\left(-1 + \exp(\mu t)\right), \quad (4.2)$$

$$\mu \equiv \alpha_b - \mu_r - \mu_a,$$

with $r$ being constant and we call $\mu$ "index number". The initial condition is

$$N_n(t=0) = N_0, \ N_b(t=0) = 0.$$

In the limit of $|\mu t| \ll 1$, we obtain the following form;

$$N_b(t) = \frac{cr}{\mu}\left(-1 + \exp(\mu t)\right),$$

$$\to crt \quad (t \to 0). \quad (4.3)$$

On the other hand, when $t \to \infty$,

$$N_b(t) = \frac{cr}{\mu}(-1 + \exp(\mu t)),$$

$$\to -\frac{cr}{\mu} = \frac{cr}{|\mu|} \quad (t \to \infty), \quad (4.4)$$

as far as $\mu = \alpha_b - \mu_r - \mu_a < 0$, where the total effect of restore and apoptosis prevails against proliferation of broken cells.

As for $N_n(t)$,

$$N_n(t) \to \left[\left(1 - \frac{N_0}{K}\right)\alpha_n N_0 - cr\right]t + N_0, \quad (4.5)$$

when $t \ll 1$. Eq. (4.5) is the same as Eq. (3.7).

## 5. Numerical Calculation of Model with Recovery Effects

We have seen that in section 4 that the time dependence of the number of normal cells, $N_n$ behaves just of the same as the one of so called logistic curve as seen in section 3, although it deviates due to the additional effects of recovery cells. While the t-dependence of the number of normal cells is not changed so drastically, the behavior of broken cells is drastically changed due to the recovery effects. Especially for the case where the proliferation rate is cancelled by the restore and apoptosis and the index number $\mu$ changes its signature and becomes negative, the number of broken cells tends to decrease while normal cells dominate the tissue. This can be clearly recognized from numerical results in Fig.4, where the vertical axis is $N$ and horizontal axis is time $t$. Solid lines correspond to $N_n$ with $\mu = -0.05, -0.1$ and $-0.5$. Dotted lines correspond to $N_b$ with $\mu = -0.05, -0.1$ and $-0.5$. The radiation rate is $cr = 10$, and $N_n(t=0)=10, N_b(t=0)=0$. From the figures, we find that the time dependence of the numbers and normal an broken cells are quite different; the one of normal cells shows almost the same behavior as the one of the case with no-recovery effects, while the one of broken cells behave quite differently if we change the index number $\mu \equiv \alpha_b - \mu_r - \mu_a$ even if we fix the irradiation strength $r$. The index number $\mu \equiv \alpha_b - \mu_r - \mu_a$ is positive when the recovery effects are very weak, and it becomes negative the recovery effects dominates.

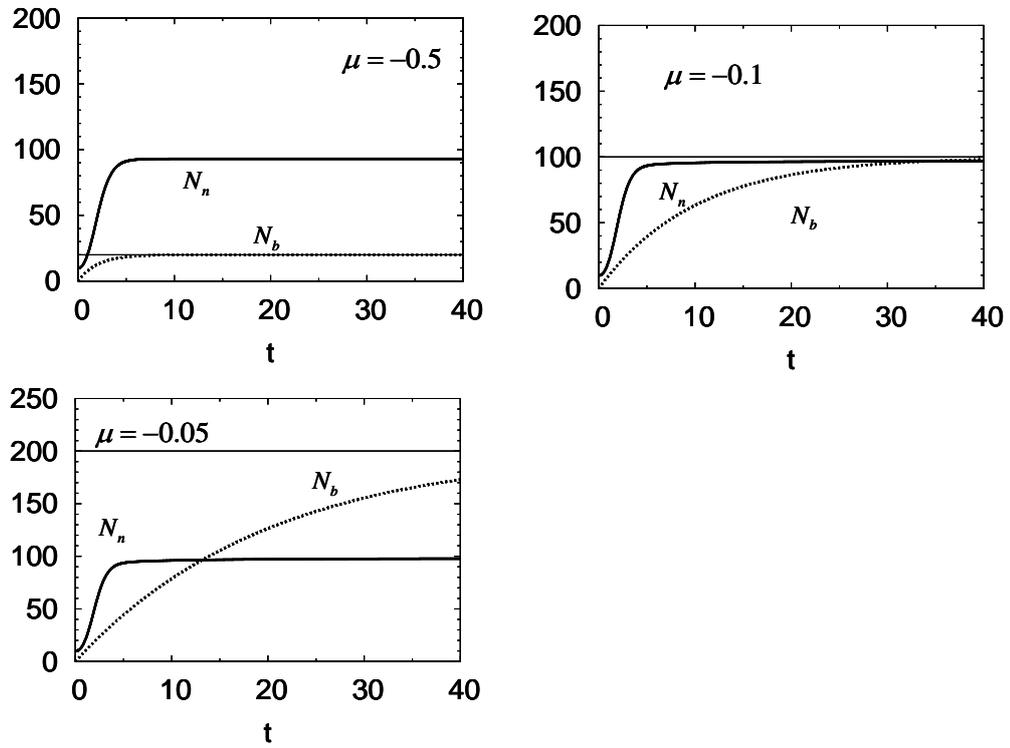

Fig. 4. Time dependence of $N_n$ and $N_b$ for various index number $\mu \equiv \alpha_b - \mu_r - \mu_a$ for fixed irradiation strength $r$. The vertical and horizontal axes express $N$ and time $t$, respectively. The solid and dotted lines correspond to time dependence of $N_n$ and $N_b$ with $\mu = -0.5, -0.1, -0.05$, respectively. The radiation rate is $cr=10$, and $N_n(t=0)=10$, $N_b(t=0)=0$, $\alpha_b=3$, $\alpha_n=1$.

From these figures, we also confirm that even in this model, the behavior is consistent with Eq. (4.3) at first stage ($t \ll 1$). Note that it is dependent only on the radiation dose rate only. Also it is to be noted that they tend to some values so far as the index number is negative: We plot those values by straight lines.

$$N_b = \frac{cr}{\mu}\left(-1 + \exp(\mu t)\right) \to crt \quad (t \to 0), \quad (5.1)$$

$$\to \frac{cr}{|\mu|} \quad (t \to \infty). \quad (5.2)$$

We have seen that the recovery effects are very important and the behavior of broken cell number is changed quite strongly so far as the index number is negative. This indicates that the lower the radiation dose rate, the safer the tissue can survive even if the total radiation dose is the same, which we will investigate in the next section.

**6. Difference between Chronic and Acute Dose Rate**

In order to see the difference between chronic and acute dose rate, we calculate the time dependence of the number of normal and broken cells for the case with the total dose $R$ kept constant by changing the time interval of radiation exposure, namely varying irradiation time $T$, with $r = R/T$ for fixed $R$. The experimental situation is as follows: first dose rate is regulated by distance and continuous exposure to a tissue is made until the total dose is accumulated up to $R$. Thus a tissue is tested, and after the time $T$, with $T \cdot r = R$, immediately it is removed from the radiation field. Such sort of experiments was done by Russel and Kelly[3] by using mice with chronic gamma ray measured mutation rate. They found a clear departure from LNT.

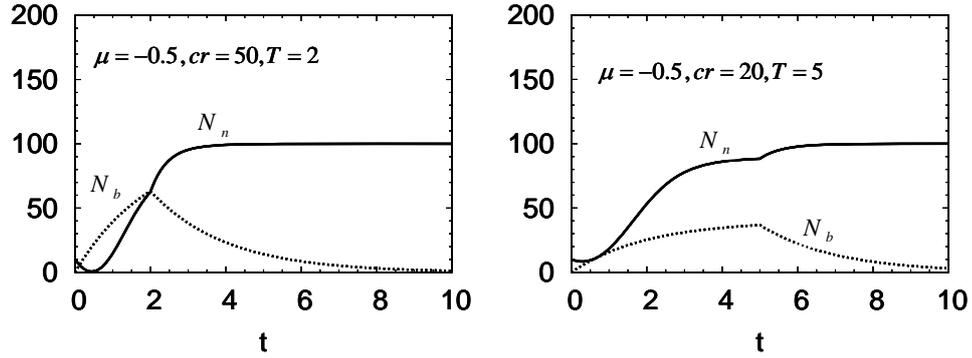

Fig. 5. Time dependence of $N_n$ and $N_b$ for various index number $\mu \equiv \alpha_b - \mu_r - \mu_a$ for fixed irradiation strength $r$ in realistic model for the case for the total dose $R = 100$ is kept constant and change the time when the radiation exposure stops after some time $T$, with $r = R/T$. As examples we show the figures when $cr = 50, 20, 10, 0.5$ with $T = 2, 5, 10, 200$, respectively, with $N_0 = 10$. The vertical and horizontal axes express $N$ and time $t$, respectively.

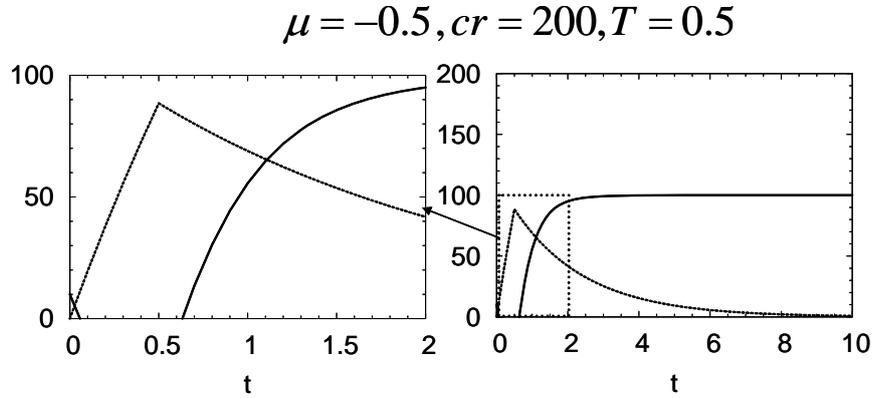

Fig. 6. Time dependence of $N_n$ and $N_b$ for various index number $\mu \equiv \alpha_b - \mu_r - \mu_a$ for fixed irradiation strength $r$ in realistic model for the case for the total dose $R = 100$ is kept constant and change the time when the radiation exposure stops after some time $T$, with $r = R/T$. As examples we show the figures when $cr = 200$ with $T = 0.5$, respectively, with $N_0 = 10$. The vertical and horizontal axes express $N$ and time $t$, respectively.

Fig. 5 and Fig. 6 illustrate numerical calculations of model with recovery effects for the case of fixed total dose $R = 100$. As examples we change the dose rate, $cr = 50, 20, 10, 0.5$ ($T = 2, 5, 10, 200$, respectively) in Fig.5 and in Fig.6 we examine more acute case, namely, $cr = 200$ ($T = 0.5$), with all the other parameters being taken the same values, $\mu = -0.5$, $\alpha = 1$.

From Fig.5 in the case of very low dose rate ($cr = 0.5$, $T = 200$), the number of broken cells is not appreciable and the tissue suffer almost no damage. On the other hand, in the case of very low dose rate ($cr = 10$, $T = 10$), the number of broken cells is appreciable however it dose not lead to canceration. If the dose rate is $cr = 20$ ($T = 5$), the number of broken cells almost approaches to the number of normal cells, however, never exceeds it. Note that at time $T = 5$ when irradiation stops, the number of broken cells begins to decrease and gradually tends to $0$. For stronger dose rate $cr = 50$ ($T = 2$), in the early stage, the number of normal cells becomes 0 and the tissue itself dies before canceration.

If we expose very acute radiation ($cr = 200$), all the normal cells will change into broken cells almost instantaneously as shown in Fig.6 and the tissue itself is driven to death, yielding serious risk of the living body. The reader may see more clearly if the detailed behavior around the period after the exposure of irradiation. Note that the behavior of the number of normal cells after some time around $t = 0.6$ where it becomes positive is actually unphysical because all the normal cells had been already disappear around $t = 0.6$. On the contrary, we recognize that the situation becomes better and if the dose rate $cr$ is lower than $20$, normal cells can survive and proliferate themselves. Thus if we stop irradiation, broken cells gradually disappear due to the restore and apoptosis effects. The time when the radiation exposure stops is indicated by the sharp change of the behavior of $N_b$. Finally the dose rate becomes very low, which may correspond to the natural radiation strength, say for example, $cr = 0.5$ as shown in Fig. 5, we can say that the living body is almost free from the irradiation damage.

## 6. Concluding Remarks

We have proposed a mathematical model which can estimate the biological risk due to exposure of radiation. Although dose rate $r$ can be time dependent and we can calculate numerically, we here concentrate on the case of constant $r$. It turns out that at starting point the number of broken cells shows linear dependence of irradiation dose rate if we switch on exposure of radiation as time goes, non-linear effect dominates, depressing its slope and finally tends to approach to the upper bound value $\frac{cr}{|\mu|}$ so long as $\mu$ is negative, which can be clearly seen in Fig.4. This indicates that, if we fixed radiation dose strength, there is some critical index number $\mu$, which is determined by the competition of recovery effects against to the proliferation, if recovery effects are stronger than proliferation of broken cells, the number of broken cells does no longer increase and tends to its maximum value. We have defined that the actual damage of a tissue occurs when the number of broken cells exceeds the one of normal cells, namely the asymmetry parameter $A$ ((Eq. (3.10)) become positive, the cancer risk becomes appreciable. Hence the condition for the asymmetry $A$ to be negative is, for large t region, is the following

$$\frac{cr}{|\mu|} < K \quad or \quad \frac{cr}{K} < |\mu|, \quad with \quad \mu \equiv \alpha_b - \mu_r - \mu_a < 0. \quad (6.1)$$

So it turns out that there is a threshold of irradiation strength rate, not total radiation dose. We should keep in mind that this condition is very important especially for long-term exposure with low-dose rate. In order to do this we have to fix the parameters and compare data so far obtained. We shall perform numerical analysis in a separate paper.

There remain many tasks to be studied. Among them we pick up some interesting problems.

First other critical point exists if the tissue is under developing phase. At $t = 0$ the number of normal cells is very small it happens the asymmetry parameter $A$ easily exceeds $0$ and the tissue turns to cancer, It is actually a little bit complicated because the number of normal cells at initial condition ($N_0$ in our notation) plays an important role.

Second is to take account of time dependence of irradiation rate. We have various types of irradiation, natural level with constant irradiation from the age of baby, X-ray exposure for checking body situation at hospital, or we have many kinds of experimental setup for biological researches.

All the above tasks shall be further investigated in near future.

**Acknowledgements**

This paper is born from intensive discussion at the meetings for investigating the risk of Low dose radiation, which has been organized by the JIFS members, without which the paper has never appeared. The authors thank to the members of LDM, especially thanks are due to K. Uno, for her valuable advices. A short comment made by T. T. Inamura[4] stimulated us to construct a mathematical model, we would like to sincere thanks to T. T.